\documentstyle[11pt,newpasp,twoside,epsf]{article}
\def\edcomment#1{\iffalse\marginpar{\raggedright\sl#1\/}\else\relax\fi}
\reversemarginpar

\begin{document}
\title{Modeling radio mini--halos in cooling flow clusters}

\author{Myriam Gitti, Gianfranco Brunetti, Giancarlo Setti, Luigina Feretti}
\affil{Dip. Astronomia Univ. di Bologna, v. Ranzani 1, 40127 Bologna, Italy\\
Ist. di Radioastronomia del CNR, v. Gobetti 101, 40129 Bologna, Italy}

\begin{abstract}
We have developed a model in which the diffuse synchrotron 
emission from radio mini--halos, observed in some cooling flow 
clusters, is due to a relic population of relativistic electrons 
reaccelerated by MHD turbulence \textit{via} Fermi--like processes. 
In this model the energetics is supplied by the cooling flow itself.
Here, the model (successfully applied to the Perseus cluster, A426)
is preliminarily applied to the possible mini-halo candidate A2626,
for which we present VLA data. 
\end{abstract}

\section{Introduction}

Several clusters of galaxies show extended ($\sim$ Mpc size) synchrotron 
emission not directly associated with the galaxies but rather diffused into 
the intracluster medium (ICM): these radio sources are called radio halos.
In some cooling flow clusters with a central dominant galaxy, the diffuse
radio emission is extended on a smaller scale, forming the so--called
\textit{mini--halos} (Feretti \& Giovannini 1996). \\
The radiative life--time of an ensemble of relativistic electrons 
losing energy by synchrotron emission and Inverse Compton (IC) scattering off 
the CMB photons is given by
$
\tau (\mbox{yr}) = 24.3/[(B^2 + B_{CMB}^2) \, \gamma]$, 
where $B$ is the magnetic field intensity (in G), $\gamma$ is the Lorentz 
factor and $B_{CMB} = 3.18 (1+z)^2 \mu$G. 
In a cooling flow region (i.e. for distances $r < r_c$, the cooling radius) 
the compression of the thermal ICM is expected to produce a significant 
increase of the strength of the frozen--in intracluster magnetic field:
$B \propto r^{-2}$ for radial compression (Soker \& Sarazin 1990)
or $B \propto r^{-0.8}$ for isotropic compression (Tribble 1993).
Therefore, in absence of a reacceleration or continuous injection mechanisms,
relativistic electrons injected at a given time in these intense fields (of 
order of some tens of $\mu$G, e.g. Ge \& Owen 1993) should already 
have lost most of their energy and the radio emission would not be observable 
for more than $\sim 10^{7\div8}$ yr.
This short lifetime contrasts with the diffuse radio emission observed
in mini--halos, hence it seems plausible that the electrons 
have been reaccelerated.

\section{Model for electron reacceleration in cooling flows}

\begin{figure}
\plottwo{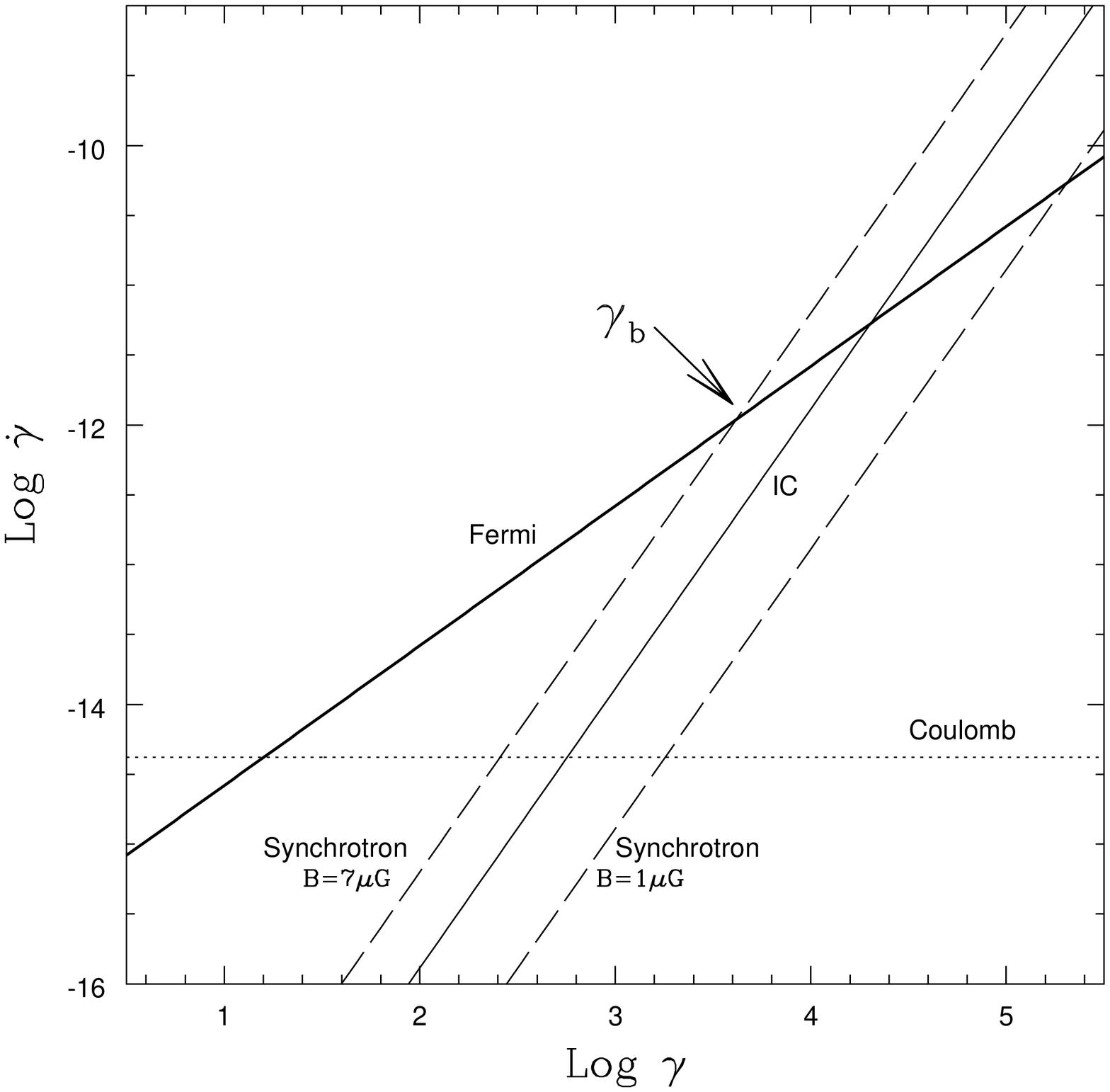}{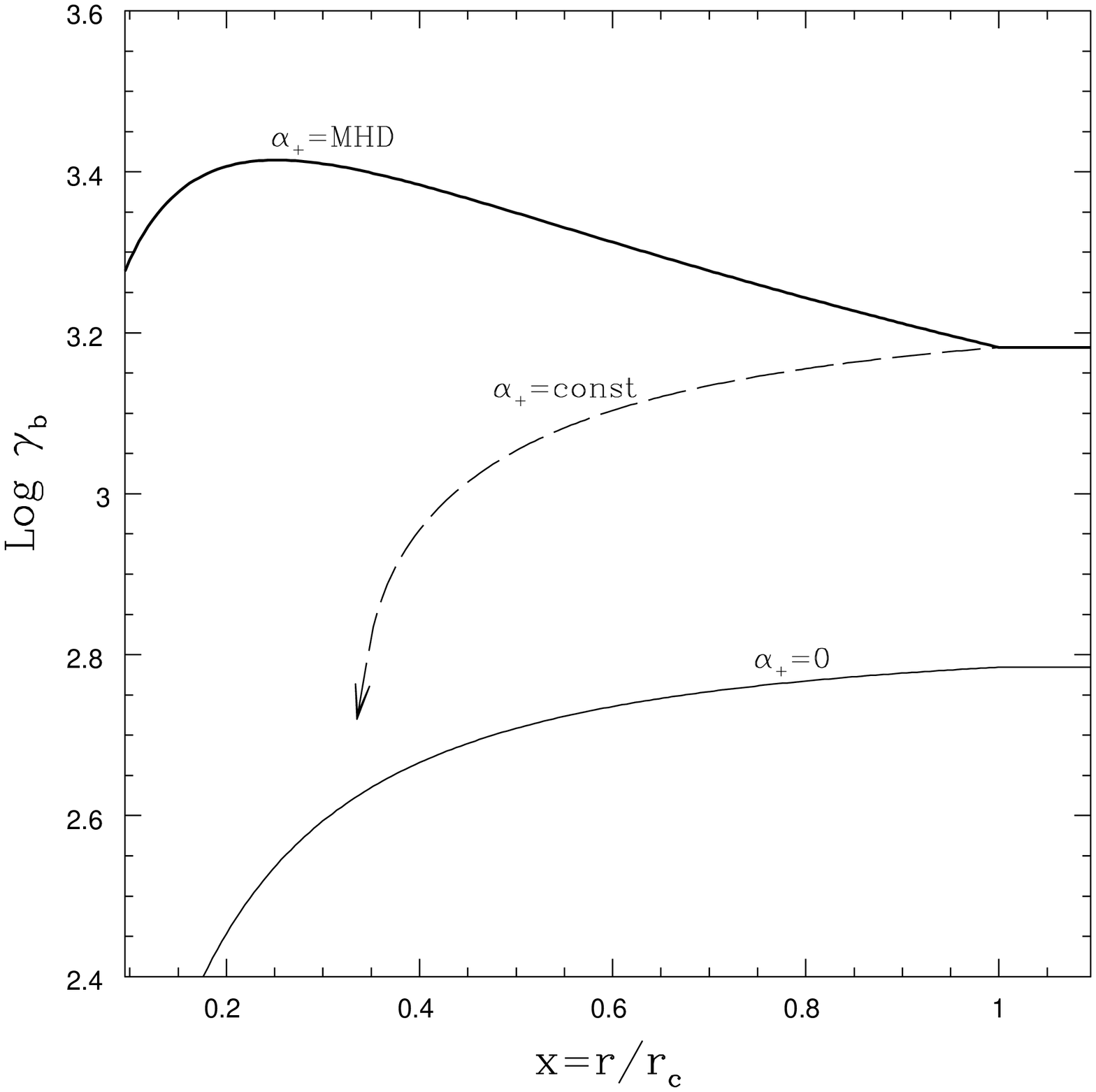}
\caption{\textbf{Left panel}: $\; \dot{\gamma}=d\gamma/dt$ is plotted for the 
processes in Eq. 1 and 2 as a function of $\gamma$. 
The value of $\gamma_b$ is shown for synchrotron losses with $B=7 \mu$G.~
\textbf{ Right panel }: Evolution of $\gamma_b$ inside the cooling flow region 
for three different cases of reacceleration (see Eq. 2): $\bullet \, 
\alpha_+$ given by Fermi--like processes related to the MHD 
turbulence in the cooling flow (bold line); 
$\bullet \, \alpha_+$=const (dashed line); $\bullet \, \alpha_+$=0, 
assuming $\Delta t \sim 4\times 10^9$ yr from injection (solid line).}
\end{figure}
The time evolution of the energy of a relativistic electron is
determined by the competing processes of losses and reaccelerations 
(both related to the magnetic field) and it is important to study the 
efficiency of these processes in the cooling flow region
as a function of $x=r/r_c$ (isotropic compression of the field is assumed): 
\begin{eqnarray}
\dot{\gamma}_{-}&=&- \beta \gamma^2 - \chi \simeq  
\, -1.3 \times 10^{-9} \left[\left(B_c^2 \, x^{-1.6} + B_{CMB}^2 \right) 
\gamma^2 + \frac{n_c}{930} \, x^{-1.2}\right] \\
\dot{\gamma}_{+}&=&+ \alpha_+ \gamma \, = f(x) \, \gamma
\end{eqnarray}
where $\beta$ is the coefficient of synchrotron and IC losses, 
$\chi$ the Coulomb losses term, $\alpha_+$ the reacceleration coefficient,
$B_c=B(r_c)$ and $n_c$ is the electron number density at $r_c$.
Since the characteristic time of radiative losses is proportional to 
$\gamma^{-1}$, while that of reacceleration is independent of $\gamma$,
the losses dominate the time evolution of the electrons for energies higher 
than $\gamma_b$, the break energy (Fig. 1, left panel).
In our model (Gitti, Brunetti \& Setti 2002) we assume that the relativistic 
electrons are continuosly reaccelerated by MHD turbulence \textit{via} 
Fermi-like processes.
This kind of reacceleration has the correct radial dependence on the 
parameters in the cooling flow region to naturally balance the 
radial behaviour of the radiative losses (Fig. 1, right panel).
Under these assumptions, the stationary spectrum of the relativistic 
electrons is:
\begin{equation}
N(\gamma) \propto (\gamma/\gamma_b)^2 \; x^{-s} \;
\exp\left(-2 \gamma/\gamma_b \right)
\end{equation}
which is essentially peaked at $\gamma_b$ and where 
we have parameterized the electron energy density as 
$\propto x^{-s}$, $s$ being a free parameter.
The other free parameters in the model are $B_c$ and $l_c$, the leading
MHD turbulence scale at $r_c$.

\section{Model results for the Perseus cluster}

\begin{figure}
\plotone{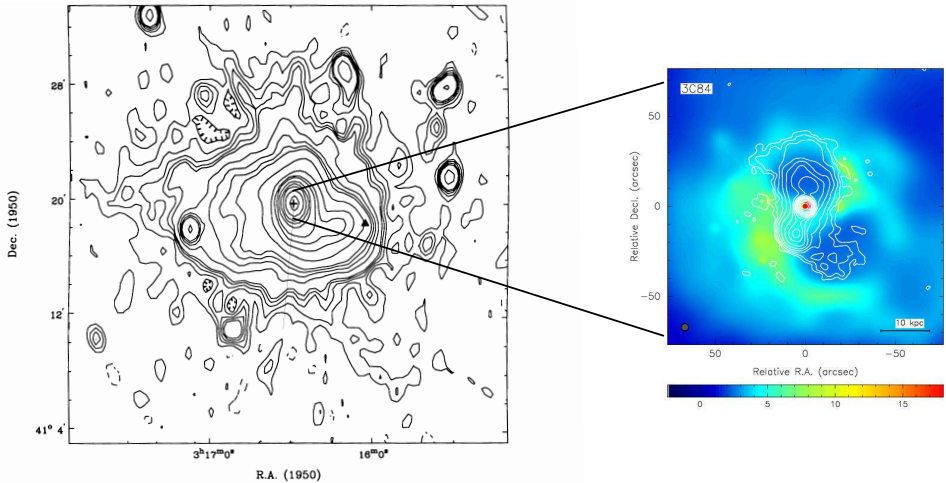}
\caption{\textbf{Left panel}: 92 cm map of the mini-halo in the Perseus 
cluster at a resolution of $51'' \times 77''$ (Sijbring 1993). 
The cross indicates the position of the cD galaxy NGC 1275. 
\textbf{~Right panel}: X--ray (grey scale)/radio (contours) overlay for the 
central part of the Perseus cluster around NCG 1275 (Fabian et al. 2000).}
\end{figure}
The first observed and well studied case of radio mini-halo is that in the 
Perseus cluster (A426, at $z=0.0183$), which hosts also a massive cooling
flow. The diffuse radio emission (see left panel in Fig. 2)
has a total extension of $\sim 15'$ (comparable with that of the cooling flow 
region, $r_c \sim 210$ kpc), a steep spectrum with spectral index 
$\alpha = 1.4$ (between 327 MHz and 609 MHz, $S_{\nu} \propto \nu^{-\alpha}$)
and its morphology is correlated with that of the cooling flow X-ray map
(Ettori et al. 1998).
On smaller scales ($\sim 1'$), B\"ohringer et al. (1993) showed evidence of 
interaction between the radio lobes of the central radiogalaxy 3C84 and the 
X--ray emitting intracluster gas. More recent results confirmed this 
interaction (Fig. 2, right panel) and support the interpretation of the 
holes in the X--ray emission as due to buoyant old radio lobes 
(Fabian et al. 2002).  
We notice that the spectral index in the lobes ranges from $\sim 0.7$
in the centre to $\sim 1.5$ in the outer regions of Fig. 2 (right panel), 
i.e. a value similar to the spectrum of the mini--halo
extended over a scale $\sim 10$ times larger. Thus
it is impossible to find a direct connection between the radio lobes and the 
mini--halo in terms of simple particle diffusion or buoyancy.
The relativistic electrons of the mini--halo should then be reaccelerated
and the morphological connection with the cooling flow
suggests a leading role of the cooling flow itself in powering the 
reacceleration process.\\
Therefore, we have applied our model to the Perseus cluster,
whose radio properties (brightness profile, integrated spectrum and 
radial spectral steepening) can be well reproduced assuming isotropic
compression of the field (Gitti et al. 2002).
In this case, the necessary energy rate to reaccelerate the relic 
population of electrons in a typical time of few $10^9$ yr is 
$\sim 3 \times 10^{42}$ erg s$^{-1}$, 
while the energy rate supplied by the cooling flow is of the order of
$10^{44}$ erg s$^{-1}$, so only a small fraction of it is required.
The number of relativistic electrons in the cooling flow region is 
$\sim 10^{62}$. This is comparable to the number of the electrons in a typical
radiogalaxy and may suggest an important role of the AGNs (and possibly of the
central AGN) in the injecton of the electron relic population.

\section{Why are radio mini-halos so rare?}

In the framework of our model, large--scale turbulence in the cluster volume 
is required to balance the radiative losses of the electrons
at $r_c$ (for Perseus we found $\gamma_b(r_c) \sim 1500$),
however at the same time the turbulence can not be too high in order to 
avoid the disruption of the cooling flow.
This scenario is similar to that of the extended radio halos where, 
however, higher energy electrons (with $\gamma_b \sim 10^4$) are required
(e.g. Brunetti et al. 2001):
this means that the turbulence efficiency would have to be correspondingly 
greater hence bringing to the disruption of a possible cooling flow,
in agreement with the observed anti-correlation between radio halos and
cooling flows.\\
Based on these considerations, we suggest that the physical conditions of the 
ICM in a cooling flow cluster with a radio mini--halo 
are intermediate between those which lead to the formation of extended radio 
halos, hosted by clusters without cooling flows, and
those holding in cooling flows clusters without radio halos.
This could explain, at least qualitatively, the rarity of radio mini--halos. 
In this scenario, where the energetics of the turbulence plays the main part 
in discriminating between the two opposite situations, 
Perseus may represent a borderline case: 
indeed there is evidence for a relatively recent merger event 
(Dupke \& Bregman 2001) which may have left a fossil turbulence, not energetic
enough to produce an extended radio halo, but sufficient for
reaccelerating the electrons producing the mini--halo.

\section{Abell 2626: a possible radio mini-halo candidate?}

\begin{figure}
\plotone{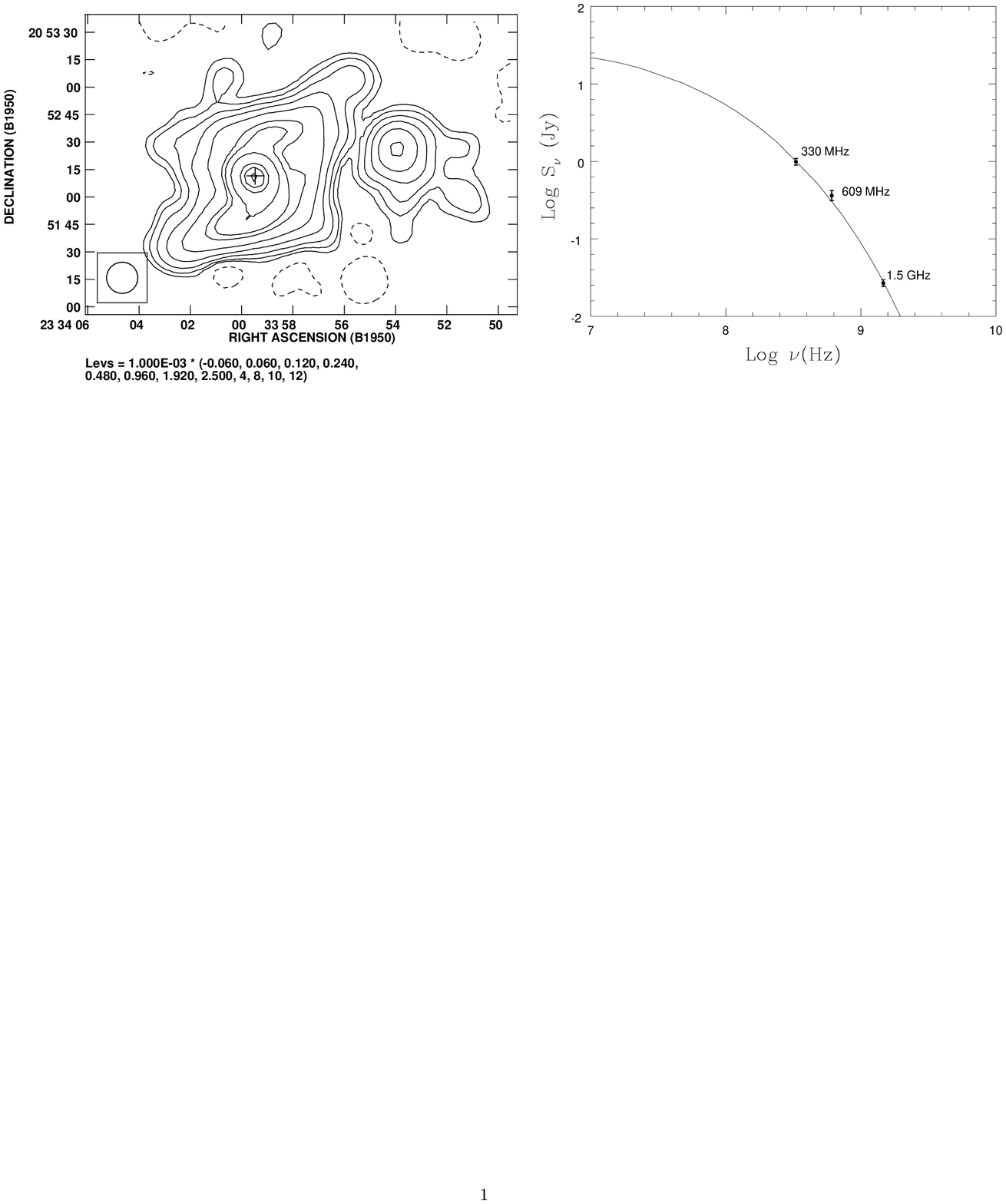}
\caption{
\textbf{Left panel}: 1.5 GHz VLA map of A2626 at a resolution of 
$17'' \times 17''$. The r.m.s. noise is 0.02 mJy/beam. The cross
indicates the position of the cluster centre.
\textbf{Right panel}: Flux densities of the diffuse emission of A2626 
compared with the total spectrum of the synchrotron emission predicted 
by our reacceleration model.
The flux density at 609 MHz is obtained by subtracting the extimated core 
emission from the total flux given by Roland et al. 1985.}
\end{figure}
In order to extend the application of our model, we have undertaken 
the analysis of A2626 ($z=0.0604$). 
This cluster has a radio emission extended on a scale comparable to that
of the cooling flow region (e.g. Rizza et al. 2000), and thus it is
one of the best mini--halo candidates.\\  
We present preliminary results for this cluster, obtained from 
VLA archive data at 1.5 GHz and 330 MHz at different resolutions.
The 1.5 GHz map (Fig. 3, left panel) shows an unresolved core (with a flux 
density $\simeq 15$ mJy, in agreement with Roland et al. 1985) and 
a diffuse emission extended by $\sim 2'$ ($S_{1.5} \simeq 27$ mJy);
we have estimated a polarization level $<2$ \%.
At 330 MHz we do not detect the emission of the core (for which we extimated
an inverted spectral index $\leq -0.6$) while the flux density of the
diffuse emission is $S_{330} \sim 1$ Jy. The spectral index 
of the diffuse emission between $\nu=330$ MHz and $\nu = 1.5$ GHz is 
$\alpha \simeq 2.4$.
These results indicate that the radio source observed in A2626
shows an amorphous structure which is likely to be not related to the central
radiogalaxy.
Thus it appears as a good candidate to apply our model for electron 
reacceleration in cooling flows. 
In Fig. 3 (right panel) we report a preliminary comparison between the
observed flux densities and the total synchrotron spectrum predicted by 
our reacceleration model.

\end{document}